# Large Language Models (LLMs) as Agents for Augmented Democracy

## Jairo F. Gudiño[1], Umberto Grandi[2], and César Hidalgo[1,3*†]

[1] Center for Collective Learning, University of Toulouse & Corvinus University of Budapest
[2] IRIT, Université Toulouse Capitole, Toulouse, France
[3] Toulouse School of Economics and Université Toulouse Capitole

## Summary

We explore an augmented democracy system built on off-the-shelf LLMs fine-tuned to augment data on citizen's preferences elicited over policies extracted from the government programs of the two main candidates of Brazil's 2022 presidential election. We use a train-test cross-validation setup to estimate the accuracy with which the LLMs predict both: a subject's individual political choices and the aggregate preferences of the full sample of participants. At the individual level, we find that LLMs predict out of sample preferences more accurately than a "bundle rule," which would assume that citizens always vote for the proposals of the candidate aligned with their self-reported political orientation. At the population level, we show that a probabilistic sample augmented by an LLM provides a more accurate estimate of the aggregate preferences of a population than the non-augmented probabilistic sample alone. Together, these results indicates that policy preference data augmented using LLMs can capture nuances that transcend party lines and represents a promising avenue of research for data augmentation.

*Author for correspondence (cesar.hidalgo@tse-fr.eu).

†Present address: Department of Social and Behavioural Sciences, Toulouse School of Economics, 1 Esplanade de l'Université, 31080 Toulouse Cedex 06 France

# Introduction

In principle, democracy is a government *of*, *for*, and *by* the people. In practice, however, democracies are constrained by trade-offs between the frequency, scope, and means of deliberation and participation. Today, many democracies rely on the use of intermediaries between the people and its sovereign power. This need for intermediation, however, has come repeatedly into question together with improvements in communication technologies.

Back in the 1960s, Joseph Licklider, an American psychologist and computer scientist, suggested that a "man-computer symbiosis" of "cooperative interaction" could perform "collaborative decision-making tasks" better than either part alone [1,2]. More recently, this need for intermediation has been questioned first, by the proponents of e-democracy and web 2.0. solutions [3–6], and more recently, by work exploring the use of artificial intelligence as a means to augment democratic participation [7–11].

In this paper, we explore the use of large language models (LLMs) as a mean to create software agents that can power augmented democracy systems. That is, we explore the use of LLMs to train personalized digital-twins that can act as intermediaries or assistants designed to augment the participatory ability of each voter [7]. We build and test different versions of this system using off-the-shelf LLMs and data collected in an experiment involving the collaborative construction of a government program during the 2022 Brazilian presidential election [12]. In that experiment, volunteers were asked to select among 67 policies extracted from the government programs of Brazil's two main presidential candidates: Luis Inácio "Lula" da Silva and Jair Bolsonaro. These volunteers were asked to select their preferred proposals out of randomly chosen pairs of proposals, providing nuanced information about their policy preferences. Volunteers were also asked to self-report a variety of demographic characteristics, including sex, political orientation, location, and age.

Here, we use this open data [12] to fine-tune six popular off-the-shelf LLMs (Llama-2 7B, LLaMA-3 8B, Chat GPT 3.5 Turbo, Mistral 7B, Falcon 7B, Gemma 7B) and explore the potential and limitations of using them to build software agents for augmented democracy. We compare the accuracy of these LLMs to the one obtained using a bundle-rule assuming that citizens with a self-reported political orientation (e.g. left/right) always select the proposals found in the government program of the candidate with the same political orientation. We find that LLM predictions tend to be more accurate than the predictions obtained from a bundle rule. This suggests that fine-tuned LLMs are able to capture nuances in a citizen's preferences that go beyond what can be predicted only from party lines. At the aggregate level, we study the ability of a probabilistic sample augmented using LLMs to predict the aggregate preferences of the population. We find that probabilistic samples augmented using LLMs provide more accurate estimates of population level aggregates of preferences than probabilistic samples alone. Finally, we introduce a diagram explaining different types of augmentation that can be applied to preference data. These findings advance our understanding of the use of software agents to create systems of augmented democracy.

## Democracy and Technology

Democracy is an institution that has long been bound and affected by communication technologies. It is hard to think about the rise of modern democracies without the printing press [13–15], just like it is hard to understand the democratic practices of the twentieth century without acknowledging the role played by newspapers, radio, and television. In the last thirty years, the communication landscape changed once again with the growth of the internet: a technology that has affected the practice modern democracy [16–18]. In this paper, our goal is not to study the impact of communication technologies on modern forms of democracy, but to explore the use of a new technology (LLMs) as a mean to construct agents to augment civic participation.

In brief, augmented democracy is the idea of using software agents to explore fine-grained forms of civic participation. These are forms that interpolate between representative and direct forms of democracy, where individuals not only choose among representatives, but can directly indicate their preferences on policy proposals [7–10].

In a representative democracy, parties or candidates represent bundles of proposals. Citizens are required to choose among competing bundles. This bundling is designed to help decrease the cognitive burden of citizens, by reducing the number of options. Yet, while there are incentives for parties and politicians to adapt their bundles to their constituents, there is no guarantee that the bundles they choose are optimal at satisfying the preferences of all citizens. By using software agents, as an alternative way to alleviate the cognitive burden of citizens, augmented democracy provides an opportunity to explore forms of civic participation that unbundle policy proposals. For instance, by allowing each citizen to train a personalized software agent that can work for them as their representative. Augmented democracy systems, therefore, could be used to estimate personalized bundles for each citizen and explore unbundled forms of democracy. That is, the use of software agents as an alternative mean to alleviate the cognitive burden of citizens is an invitation to explore the creation of collective decision-making systems that could be hard to build in the absence of this technology.

LLMs provide an interesting opportunity for the design of augmented democracy systems as they satisfy a few important conditions: (i) they are easy to use, (ii) they operate directly over natural languages, and (iii) they are part of a competitive market populated by a wide variety of suppliers. LLMs language abilities make them an interesting choice for the creation of systems interacting directly with policy proposals and citizen preferences expressed as text. In fact, LLMs are good at simulating human responses in opinion polls [19–22] or predicting votes in binary elections [19]. Also, since LLMs are trained on large bodies of text, they are likely to encode information on the policy preferences of a wide variety of people, which can be potentially extracted with the right prompting (e.g. by using backstories to create personas [19]) or fine-tuning. From the perspective of industrial organization, today LLMs are part of a competitive global market including hundreds of options (the latest Open LLM leaderboard on Hugging Face contains hundreds of LLMs [23]). This competitive market is an important institution, as the ability of people to switch among LLM providers reduces the risk of manipulation and/or capture. Yet, there are also important caveats that we need to consider when exploring the augmented democracy potential of LLMs.

LLMs are known to exhibit biases and limitations [24–27], which could set a ceiling on their ability to represent people with different political views and opinions, especially those groups that do not participate in the creation of potential training data (e.g. remote and offline indigenous communities). Also, LLMs are a powerful technology following a wide variety of governance models which can lead to different forms of manipulation and capture. Some LLMs, such as those released by Open AI, are developed as proprietary models in ways that are untransparent about their source of training data. Other LLMs, such as LLaMA, are developed in an open-source model but still rely heavily on the support of researchers employed in a private sector organization. Other LLMs such as Mistral, rely on venture funding, while others, such as Falcon, depend directly on government support, in this particular case, the support from the government of Abu Dhabi. All of these governance models are not immune to potential capture or manipulation. Finally, LLMs exhibit poor logical skills in some tasks (they have been notoriously bad at math and struggle with spatial reasoning), which could be problematic in problems requiring these skills. These limitations need to be taken seriously before any real-world implementation of an augmented democracy system.

Our work also complements several studies exploring different aspects of digital and/or augmented democracy. The technical literature on augmented democracy includes efforts focused on cryptographic solutions for privacy and verification [28,29] as well as work on data augmentation using matrix completion techniques [29] or LLMs, as shown in recent work on participatory budgets [8] or direct democracy in Switzerland [30]. This technical work has also explored the creation of deliberative forms of augmented democracy—where LLMs respond to each other—by creating conversational social media agents [31] or by using LLMs to simulate the responses of citizens with different political views [19,32]. The idea of augmented democracy builds also on recent work showing that LLMs can be used to simulate human participants in surveys [19–22], which has shown for instance, that LLMs provide similar moral judgements than humans [20] and can be used to construct fine-grained personas and predict their electoral behavior [19]. On the philosophical side, the exploration of LLMs has focused mostly on the critical and ethical comparison of different forms of digital democracy, including augmented democracy [9,10,33,34,34,35].

Finally, it is worth noting that augmented democracy is an idea that stems naturally from recent advances in crowdsourcing and artificial intelligence [12,36–43]. The setup used in this study closely resembles work in urban planning, where a paired comparison system was used to collect information on people's preferences of streetscapes [39], and then used to train machine learning models that augmented that data to produced fine-grained evaluative maps of cities [44–46]. In this paper, we follow a similar setup, but instead of using information about people's preferences over streetscapes, we use information on their preferences across policy proposals. We use this setup to introduce different modes of augmentation and to develop benchmarks for both, individual and aggregate preferences. These findings contribute to our understanding of the use of LLMs as a mean to create agents for augmented democracy.

## Results and Methods

Figure 1 a and b shows the basic data we use to train and evaluate LLMs predictions. This is data collected in Brazucracia.org, an online participatory experiment conducted during Brazil's 2022 presidential election [12]. In Brazucracia participants were asked to select among pairs of proposals extracted from the government programs of the two main candidates of the 2022 Brazilian election (see Figure 1a). For instance, to prioritize among: *"investing in clear and renewable energy,"* a proposal that was present in the program of both Lula and Bolsonaro, or *"strengthening of the subsided pharmacies program,"* a proposal that was present only in the government program of Lula.

During the process, participants were also invited to fill out a basic demographic survey (age, political ideology, rural/urban area, educational attainment, gender, age, and geographic location), which we can use to connect their preferences to their stated demographic characteristics. The collection of this data was approved by TSE-IAST Review Board for Ethical Standards in Research, under the reference code 2022-07-001 and was made publicly available together with the publication of [12]. Our dataset consists of 8,719 pairwise preferences elicited by 267 participants over a universe of 67 proposals. While this data is sparse (which is one of the conditions that motivates the need for augmented democracy), we estimate the test-retest reliability of the aggregate preferences of the full sample at 95.38% (see supplementary material section B). This is consistent with previous work using paired comparison data in other contexts[39], since paired comparison rankings tend to converge at about 30 to 40 preferences per participant. More details about the dataset and data adequacy checks are presented in Appendices A and B of the Supplementary Material and in reference [12].

We use this data to explore the ability of the LLMs to model the individual and collective preferences of this population of 267 participants. Going forward this population represents our statistical universe and will refer to it as the full sample or complete sample of participants. That is, this study is not focused on estimating preferences for the general population of Brazil (as we lack the data to do so), but on understanding how samples from this universe of participants (e.g. a 20% random sample of only 53 participants) can be used to estimate the preferences of the entire universe of participants (the 267 participants).

First, we will explore the ability of LLMs to model individual preferences. That is, we will use a train-test cross-validation setup to study the ability of the LLMs to predict the preferences of individuals withheld from the training data but available in the test data. Then, we will study the overall ability of the LLMs to reproduce the aggregate preferences observed in the data, by comparing the ability of pure and augmented random samples to reproduce the aggregate preferences of the full population of participants.

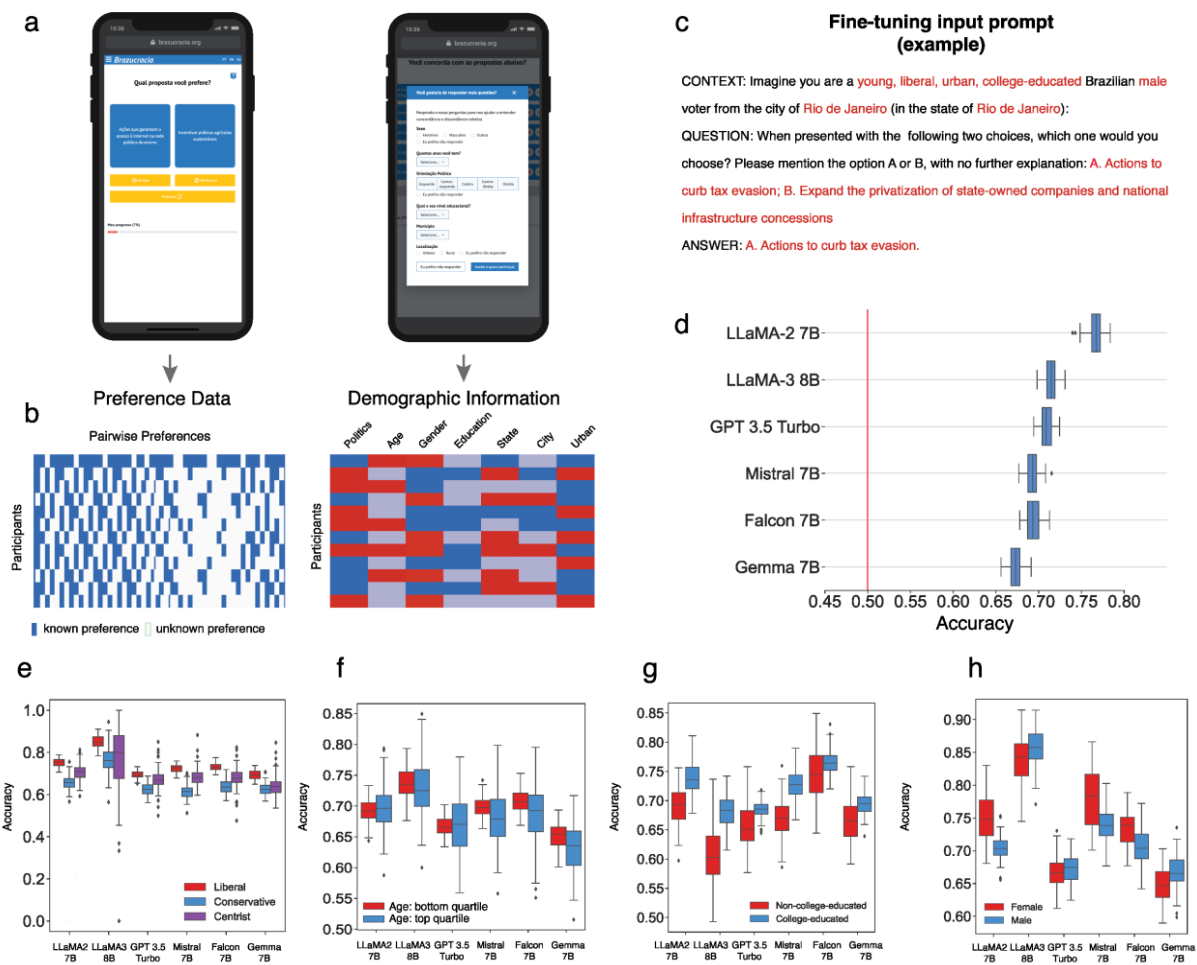

**Figure 1 a** Data was collected in brazucracia.org, a collaborative government program builder deployed during the 2022 Brazilian presidential elections [12]. **b** Brazucracia data consisted of pairwise preference data and demographic data for each participant. **c** Example of the "mad-libs" style prompt used to fine-tune the LLMs. **d** Accuracies obtained on the 50% test set of six LLMs. **e–h** Comparison of the accuracies obtained for samples with different self-reported demographic characteristics, by **e** political ideology, **f** age (older and younger quartiles), **g** education, and **h** sex. Samples in figures e-h were balanced by downsampling the variable with the largest representation in the data. Confidence intervals at 99% are calculated through bootstrapping with 100 iterations.

We begin by splitting our data randomly into a training set (50% of participants) and a test set (50% of participants). We use the training set to train LLMs using the following procedure (Figure 1 c):

First, we create prompts that encapsulate the demographic characteristics of each participant. For example, for a 26 years-old liberal male with a master's degree living in Rio de Janeiro, who in *Brazucracia* selected the proposal: "*actions to curb tax evasion*" over the proposal "*expand the privatization of state-owned companies and national infrastructure concessions.*" We then feed the LLM such prompts as well as a prompt reversing the order of the proposals to compensate for the fact that LLMs exhibit biases depending on the order of the text [27] (Figure 1 c).

We use these prompts to fine-tune LLMs by retraining a fraction of their parameters (~12%) using a Low-Ranking adaptation (LoRA) technique, a method used in NLP to improve the performance of pre-trained LLMs on predefined tasks [47]. LoRA is a Parameter-Efficient Fine Tuning (PEFT) method that works by freezing the model weights of the network and efficiently recalculating a fraction of the weights of the network to adapt it to the injected prompts. This allows the creation of customized LLMs with relatively little parameter adjustment. For reproducibility reasons, we fine-tune at temperature zero (but find that using different temperature parameters does not affect our results (see SM Section H)).

Next, we test the ability of LLMs to predict the preferences of participants using six off-the-shelf LLMs: (i) GPT 3.5 Turbo [48], a model trained with proprietary architecture and data, (ii) LLaMA-2 7B, (iii) LLaMA-3 8B [49] (iv) Mistral 7B [50] and (v) Gemma 7B [51], four open-weight models trained on proprietary data, and (vi) Falcon 7B [52], an open-weight model developed by a government sponsored research laboratory trained on public data. We also include a foundation model trained on public data (BERT SQuAD - BERT for Multiple Choice) [53] as a benchmark. We use this simpler model (of an older vintage and with fewer parameters) to test whether present day LLMs, which are oversized and more intensive on their use of computational resources [54], are needed for this augmentation task or if a simpler model would do. We do not present results for GPT-4 and Claude 3 as fine-tuning options were not yet publicly available at the time of writing this paper. Details of the fine-tuning process for each LLM are presented in Appendix C of the Supplementary Material.

## Individual Preferences

Figure 1 d shows the percentage of times that a fine-tuned LLM trained on a random sample of 50% of the participants correctly predicts a pairwise choice from a participant in the test-set composed of the remaining 50%. Ninety five percent confidence intervals are calculated through bootstrapping with 100 iterations. In this test, LLaMA-2 registers the highest accuracy (76.68% ± 0.0014) followed by GPT 3.5 Turbo (70.4% ± 0.0013), Mistral 7B (69.4% ± 0.0017), and Falcon 7B (69.3% ± 0.0014). BERT SQuAD does not perform better than random (50.83% ± 0.0015) (see SM Appendix E), making it relatively unsuitable for this data augmentation task. In SM Appendix E we present results for fine-tuned LLMs trained on random samples of 5%, 25% and 75%. In SM Appendix F we present results for the Kendall's tau as an alternative metric of accuracy (which compares the number of congruent pairs (when the LLM and the citizen made the same choice) and incongruent pairs. Our data cannot be used to estimate an F1 statistic, since by construction, it couples true positives and true negatives, and false positives and false negatives, making F1=Precision=Recall=Accuracy.[*]

Next, we explore how the accuracy of these models relates to the demographic characteristics of the population of participants. This will help us explore the potential biases of these models. For instance, we would like to determine if LLMs predict better the preferences of college-educated participants by comparing the accuracies

---

[*] We do not present an F1 score is because our data structurally couples true positives (TP) and true negatives (TN) and false positives (FP) and negatives (FN). Consider an LLM that is asked to make a choice between options A and B in a dataset where we know the user chose option A. If the LLM chooses A we have a true positive for A and a true negative for B. If the LLM chooses B we have a false negative for A and a false positive for B. Hence precision:

Precision = TP/(TP+FP) and Recall = TP/(TP+FN) are the same

Since FP=FN (Precision=Recall)

Thus, F1 = 2TP/(2TP+2FP) = TP/(TP+FP) = Precision = Recall = Accuracy.

obtained when predicting the preferences elicited by college and non-college educated participants in the test set.

Because our data is unbalanced (e.g. we have more liberal than conservative participants), we retrain our models using balanced data subsamples generated by randomly selecting a smaller sample from the overrepresented subset. For instance, if our training data contains 70 individuals associated with characteristic A and 30 associated with characteristics B, we fine-tune a new model using the 30 individuals associated with B and a random sample of 30 individuals associated with A. We then generate predictions for individuals in the test set using these LLMs and compare the accuracies obtained for individuals associated with A and B.

Figure 1 e to h show the accuracies obtained by the LLMs on subsets of participants with different self-reported political views (liberal/conservative) (figure 1 e), age (younger and older quartiles) (figure 1 f), education (college educated/non-college educated) (figure 1 g), and sex (male/female) (figure 1 h).

Across all six LLMs we find accuracies to be higher when predicting the preferences of liberal participants compared to conservatives and centrists (all p-values<0.01 see SM Appendix I). When it comes to age, we find a slight but significant tendency to predict the preferences of younger participants more accurately in Mistral 7B, Falcon 7B and Gemma 7B. When it comes to education, we find that all LLMs predict the preferences of college educated participants more accurately than those of non-college educated participants (all p-values < 0.01, see SM appendix I). Finally, when we split our data by self-reported sex, we find a mixed bias. While LLaMA-2 7B, Mistral 7B and Falcon 7B are better at predicting the preferences of females, LLaMA-3 8B and Gemma 7B are more accurate at predicting the preferences of males. This contributes to the ongoing discussing on whether LLMs may overrepresent some segments of the population [55,56].

Next, we benchmark the accuracy of the LLMs against a bundle rule, representing the idea that voters in a representative democracy are required to choose among *bundles* of policies represented by parties or politicians. A bundle rule prediction consists of choosing proposals listed on the program of the candidate that matches the political ideology of each participant. That is, predicting that a self-reported left-wing liberal chooses a proposal listed in Lula's program and a self-reported conservative chooses a proposal listed in Bolsonaro's program. We test the bundle rule benchmark using two different exercises.

First, we compare the accuracy of different LLMs disaggregated by the ideology of the participant and the ideology of the proposals. More concretely, we compare these accuracies using matrices that calculate the percentage of times that a preference is predicted correctly when: (a) a liberal chooses a policy listed in Lula's program (top-left) against a proposal listed in any other program (but not in Lula's); (b) a liberal chooses a policy listed in Bolsonaro's program (top-right) against a proposal listed in any other program (but not in Bolsonaro's); (c) a conservative chooses a policy listed in Lula's program (bottom-left) against a proposal listed in any other program (but not in Lula's); (d) a conservative chooses a policy listed in Bolsonaro's program (bottom-right) against a proposal listed in any other program (but not in Bolsonaro's). We exclude from the exercise cases in which the participant chooses between two proposals coming from the same candidate (e.g. a choice between two proposals present only in Lula's program) but explore this case later.

Figure 2 a show that, across most cases, LLMs exhibit higher accuracies than the bundle rule (21 out of 24 cases or 87.5% of times). These differences are sometimes large. LLaMA-2 7B is 15 percentage points more accurate than the bundle rule at predicting which proposals from Lula are selected by self-identified left-wing participants. In fact, all LLMs get an accuracy of 80% or higher when predicting the policies of Lula's program selected by self-identified left-wing participants (compared to 71% for the bundle rule). Yet, we do find three exceptions, involving liberals choosing a policy listed in Bolsonaro's program for Mistral, Falcon and Gemma.

Our second approach, shown in Figures 2 b and c, considers predictions that cannot be made using the bundle rule. These are predictions involving a choice among proposals from the same candidate (e.g. preferences among two proposals from Lula's program). In this case, the LLMs still have significant predictive power for these cases (between 65% and 77%). Overall, we find that LoRA fine-tuned LLMs are better at predicting individual preferences than what we get from predictions based solely on self-reported political orientation.

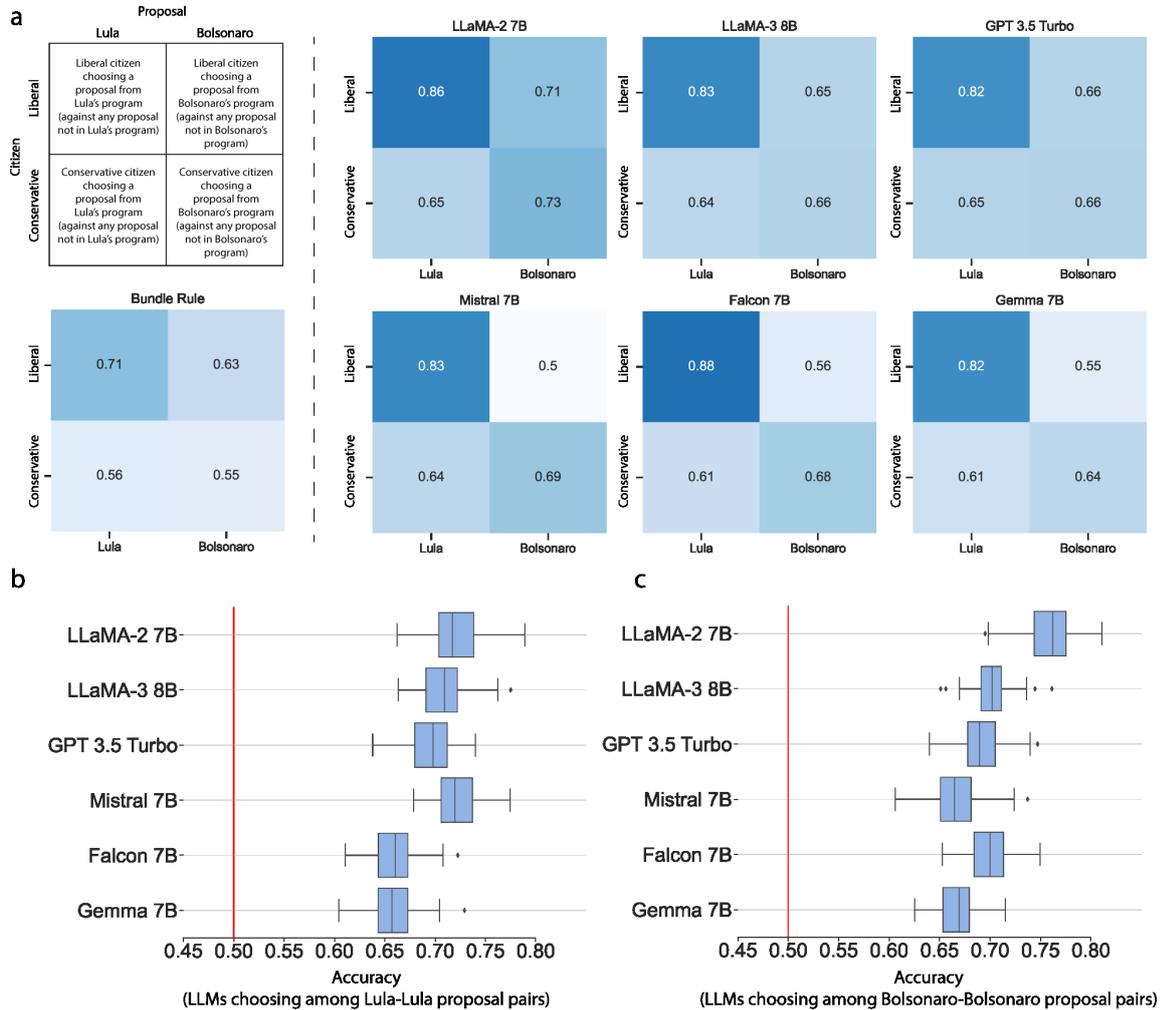

**Figure 2** comparison between LLMs and the bundle rule. **a.** Accuracies obtained for the six different LLMs and the bundle rule when predicting pairwise preferences. **b-c** Accuracy of LLMs when predicting preferences between pairs of proposals found in the same program, **b** Lula, and **c** Bolsonaro.

## Aggregate preferences

Next, we explore whether we can use these LLMs to improve our ability to estimate the aggregate preferences of the population starting form a random sample. As a benchmark, we estimate the accuracy with which a probabilistic sample of participants predicts the preferences of the full sample. That is, we ask if a probabilistic sample plus an LLM trained on that same sample is better at predicting the aggregate preferences of the full population than the probabilistic sample alone.

As a measure of aggregate preferences, we estimate the win-rate of each policy proposal by aggregating the data obtained from the individual predictions. This is a Borda inspired score that we can estimate for incomplete preference data. It is defined as the fraction of times a proposal is selected out of a pair. A win rate of one (or 100%) indicates that a proposal was always chosen over others and a win rate of zero indicates that a proposal was never chosen over others. Win rate is, therefore, a measure of the overall acceptance of a proposal among the population of participants.

Formally, let $w_{ij}$ be the number of times proposal $i$ was selected over proposal $j$. Then the win rate $W_i$ of proposal $i$ is defined as:

$$W_i = \frac{\sum_j w_{ij}}{\sum_j (w_{ij} + w_{ji})}$$

To estimate the ability of a random sample to represent the aggregate preferences of the full population we need to estimate the win rate $W_i$ of each proposal twice, once for the full sample (all 267 participants and all their elicited preferences) and another time using data from a random sample (Figure 3 a). We then compare the similarity between the estimated win rates by calculating the $R^2$ of their Pearson correlation (Figure 3 b and c). An $R^2$=100% indicates that the win rates obtained using the partial sample are identical than those obtained for the full population, meaning that the sample can reproduce the aggregate preferences of the larger population.

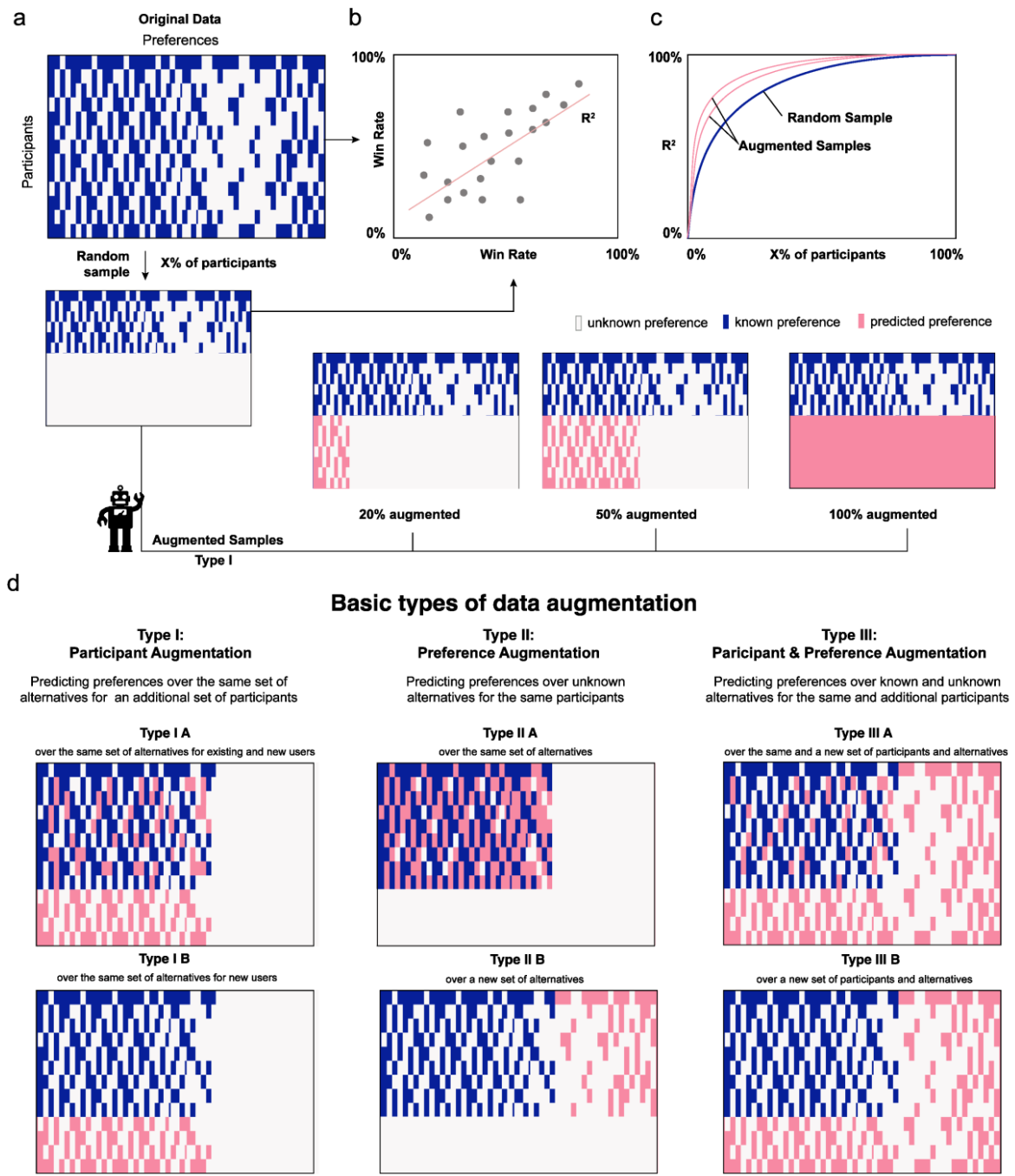

**Figure 3** augmentation and validation procedure. **a** We represent preference data using a matrix in which each row represents a participant and each column a preference over a pair of policy proposals. We can sample this data randomly, and augment that sample, to estimate the ability of the random and augmented samples to reproduce the data of the complete sample. **b** We assess the ability of a random or an augmented sample to reproduce the aggregate preferences of the full sample by comparing the win rates among the proposals and estimating the resulting $R^2$ statistics. **c** We plot the $R^2$ of the win-rates comparing a random or an augmented sample of size x with the full sample as a way to estimate the accuracy with which that sample of size x represents the aggregate preferences of the full sample. **d** Different data augmentation procedures.

Next, we explain our data augmentation procedure (Figure 3 d). In principle there are multiple ways in which one could augment preference data. Figure 3 d shows three types of data augmentation (Types I to III). Type I involves augmentation of the participants in the sample, which involves predicting the preferences of additional participants based on external information about them, in our case, demographic characteristics (e.g. age, sex, education, etc.). Type II involves predicting additional preferences for the existing set of participants. This could involve predicting unelicited preferences over the same alternatives (Type II A) or predicting preferences for new alternatives (Type II B). In our case, Type II B would include predictions over alternatives that were not part of the 67 proposals presented in Brazucracia. There is also a Type II C which combines these two. Finally, Type III involves predictions for both, new participants, and new alternatives (and is a combination of Types I and II).

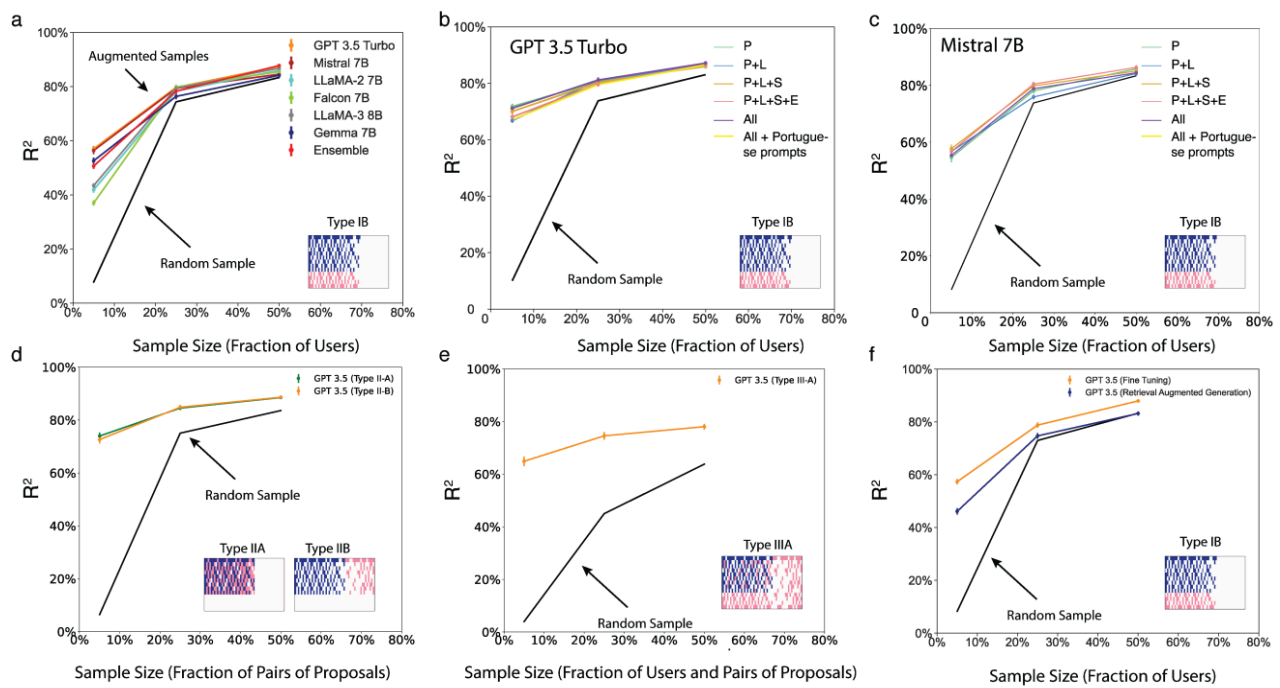

**Figure 4.** Comparing augmented and non-augmented random samples. We estimate the accuracy with which a sample of size x% is able to reproduce the aggregate ranking of preferences of the full population using the $R^2$ statistic. **a.** Comparison between a probabilistic sample of size x percent of users and an augmented sample (Type I B) generated with LLMs fine-tuned on the same data. **b.** Same as **a** but using a different set of demographic characteristics to train each LLM (P=Political Ideology (liberal, conservative, centrist), L=Location (city and state), S=Sex (male, female), E=Education level (non-college educated, college educated). **c** Same as **b** but for Mistral 7B. **d.** Type IIA and IIB augmentations using all demographic information. **e.** Type IIIA augmentation using all demographic information. **f.** Comparison of augmentation via fine-tuning and RAG for Type IB augmentation.

Figure 4 a show a direct example of a Type I B augmentation process using Chat GPT 3.5 Turbo, LLaMA-2 7B, Falcon 7B, Mistral 7B, LLaMA-3 8B, and Gemma 7B. The black dashed line shows the ability of a non-augmented random sample of 5%, 25% and 50% of participants to reproduce the aggregate preferences of the full population and the colored lines show the accuracy of this data augmented by each of these LLMs. We use this LLM to augment the data by predicting the elicited preferences of an extra 20% of the remaining population (e.g. 20% of the remaining 95% of the population in the case of the 5% sample). For the 5% random sample this provides a substantial boost, from about $R^2 \sim 30\%$ to $R^2 \sim 75\%$, and even at a 25% random sample the boost provided by the LLMs is substantial (about 7 to 10 percentage points). We do not provide results for non-fine-tuned models as these models always choose the first alternative (A) at zero temperature. Finally, we present a consensus method (ensemble) that picks the prediction made by the majority of the LLMs but find that this does not work better than the single LLMs. These results show that an augmented sample can be better at reproducing the preferences of the full population of participants than a random sample alone.

Figure 4 b and c repeat this procedure while varying the demographic characteristics used to train the LLMs (e.g. including or not information about a participant's level of education) for both GPT 3.5 Turbo and Mistral 7B. Sensitivity tests for temperature are presented in Appendix H. Overall, we find that the accuracy of the augmented samples does not depend strongly on the demographic variables included or if we change the language of the training prompts and the consultation from English to Portuguese [57]. In SM Appendix D we present the prompt and description of features in Portuguese.

Figure 4 d shows examples of Type II A and B augmentation using Chat GPT 3.5 Turbo. In this case, the augmentation process is even stronger than for Type IB. We conjecture this is due to the fact that Types IIA and IIB make better use of preference data, since each user is not characterized only by their demographic characteristics, such as in Type IIB, but also by a sample of its elicited preferences. Figure 3 e show akin results for a Type III A augmentation.

But what is the knowledge captured by these LLMs? Does it go beyond simple context? To explore this questions we compare two prediction methods implemented in GPT 3.5 Turbo: (1) fine-tuning with LoRA ([47]) and (2) RAG (Retrieval Augmented Generation [58]), a method based on providing additional context in the form of documents to the LLMs, in our case a document summarizing the proposals of the participant's favorite candidate (Bolsonaro for conservatives; Lula for liberals). This comparison helps us evaluate whether the performance of the fine-tuned model comes from its ability to capture information that goes beyond the context made available through the RAG approach. We find that the LoRA fine-tuned predictions reproduce the preferences of the full population of participants better than RAG (Figure 4 f), especially for larger samples (25% and 50%). This validates the idea that the fine-tuning process can be used to create personalized software agents that provide a more accurate representation of an individual's preferences than software agents created by providing context to the chat layer of the LLMs.

# Discussion

Is the world ready to explore augmented forms of civic participation? In 2019, IE's University Center for Governance and Change published a report where 2,576 people from eight European countries responded to questions about digital technology [59]. According to Calvo and Garcia-Marza [60], this revealed a digital paradox, since 70% of respondents believed "digital transformation needs to be controlled to avoid" its negative impacts in society, while a non-negligible proportion of the respondents (25%) indicated they be "in favor of letting an artificial intelligence make important decisions about the running of their country." More recent studies also provide some credence to the idea that people may be willing to consider using AI in policy making. A study published in 2024 [61] reported that more than 60% of the people in their sample would approve of a decision making model in which AI has 25% of the decision making weight and politicians 75%. The same study showed that people also are more willing to give some weight to an AI in technical tasks, but still would prefer to defer to experts in that case.

These numbers provide some food for thought. On the one hand, we live in a world that is anxious about the societal implications of digital technology. On the other hand, there seems to be an important number of citizens that is willing to give some civic power to AI. But are those willing to give AI a chance in theory, are willing to do so in practice? The evidence so far points to the contrary. A first wave of efforts to create centralized AI representatives such as SAM in New Zealand [62], Alisa in Russia [63], Leader Lars in Denmark [64], ION in Romania[65], and AI Mayor in Japan [66], gathered more media support than success at the ballot box. The case of AI mayor in Tama City, a 150,000 people suburb of Tokyo, is particularly interesting, since the support for the AI mayor decreased instead of increased in subsequent elections. In its first appearance, in 2018, it made it to the second round after receiving 4,013 votes. In a more recent election in 2022 AI Mayor received only 516.

Against this backdrop of efforts, augmented democracy remains still a relatively unexplored idea [8,30]. It is also different from the creation of AI politicians, as it does not involve the creation of a single AI representative designed to "listen to everyone," but an ensemble of AI agents, each controlled by its own human: citizens can create individual profiles that can be personalized according to their own characteristics, preferences and habits, and these autonomous agents can potentially vote on their behalf.

In this paper, we contributed to the early exploration of augmented democracy systems by fine-tuning six off-the-shelf LLMs and studying their ability to anticipate organically collected fine-grained political preference data collected during the 2022 Brazilian presidential election [12]. We found that LLMs predict pairwise preferences with higher accuracies for participants that self-reported as younger, liberal, and more educated. We then explored the ability of LLMs to augment participation data in an exercise in which we used a sample of our data to predict the aggregate preferences of the full population of participants. This exercise showed that the LLMs provide an effective augmentation for small sample sizes (<30%), resulting in estimates of the aggregate preferences of a population that are more accurate than those obtained from non-augmented random samples.

We also found the LLMs to be more accurate than a bundling rule (assuming people always select the policies of their preferred party or politician) suggesting that LLMs capture information that is more nuanced than a simple left-right wing divide.

Yet, despite these findings, this paper still leaves many unexplored questions. On the one hand, we use LLMs only in a context of preference aggregation, where the idea of using AI to explore augmented forms of multi-agent consensus is also a promising avenue of research [67–69]. Also, we do not explore how traditional forms of multi-agent consensus might change in the context of preferences being elicited in an augmented democracy system. On the other hand, future research must explore whether the performance of LLMs is contingent on the design of the platform. Our results are based on data from Brazucracia, a platform that differs from the design of other Voting Advice Applications (like Wahl-O-Mat or Elyze) in an important aspect. In Brazucracia, the objective is to rank-sort the most relevant proposals for citizens whose preferences might not align perfectly with a specific political party, instead of recommending a party or politician to the user. We also do not use multiagent systems, which in some cases have been shown to be more accurate than single agent LLMs [70,71]. Finally, our exploration of LLMs was far from comprehensive, and did not include some of the latest models (e.g. GPT-4 and Claude 3) or multiple variations in the prompts used to train and query the LLMs. But these are only some of the limitations of our work.

There are also important limitations involving the representation of both, preferences, and participants. The idea of augmented democracy is based on the construction of digital twins, which in the case of this paper, were constructed using a minimalistic representation of each agent (a relatively short vector or demographic characteristics and pairwise preferences). This is a far cry from the state of the art of digital twin creation, such as the recent digital twin demo released by LinkedIn founder's Reid Hoffman [72]. The demo involved a video interview in which Hoffman interviewed and was interviewed by his digital twin, which was trained among other things on many of his books. Certainly, digital twins could be made more accurate with more and better data, but the differential availability of that data adds to the challenge of political representation, since the production of the data needed to train AI systems (e.g. text, video) is uneven among the population.

Today, LLMs are not ready for deployment in full-fledged augmented democracy systems but provide an interesting avenue of research in that direction. This research needs to address key concerns.

First, the data used to pre-train many of these LLMs (e.g. GPT 3.5 Turbo and Mistral 7B) is proprietary and could be open to manipulation. The sensitive nature of augmented democracy system requires us to think deeply about the open-source code and open data rules needed to develop these systems in a transparent manner.

Second, LLMs can generate ambiguous predictions that sometimes depend on the order in which options are presented (A vs B, or B vs A). In some cases, this consistency can be lower than 70% (see SM Section G).

Third, LLMs exhibit biases. In this paper, we found LLMs to be more accurate at predicting participants that self-reported as liberal and more educated, and some LLMs tended to exhibit a bias in favor of females. This talks to the long literature in NLP and LLMs discussing gender bias [73–76]. Yet, our results are somehow different since that literature focuses largely on how LLMs use language, for instance, to disambiguate words such as *doctor* and *nurse* (e.g. assume doctors are males and nurses are females). In this paper, we are not using LLMs to disambiguate gender but to predict their civic preferences. Thus, we cannot assume that the same bias (e.g. favoring males) operates in this context. Still, there are some important points of comparison in the recent literature.

A recent paper by Argyle et al. [19] uses LLMs (GPT3) to simulate human responses in surveys. Argyle finds that LLMs given personal backstories constructed to represent a particular demographic are able "*to accurately emulate response distributions from a wide variety of human subgroups.*" In their study, Argyle et al. compare voting predictions made by these LLMs by gender, finding the LLMs to be slightly more accurate at predicting the preference of females (about one percentage point). Santurkar et al.[21] is another interesting study that also uses LLMs to explore human opinions in a question/answering setup. In their case they find LLMs tend to more accurately reproduce the preferences of young, low-income, moderates with less than high-school education. Santurkar et al. [21] reports that the LLMs used in their study are more accurate at predicting the preferences of males, but these differences are also small (about one percentage point or less).

Fourth, we lack a good framework of explainability, since we do not have a thorough understanding of *why* LLMs choose one proposal over another.

These are a few of the many limitations involved in an approach like the one we present here, which means that much work remains to be done. Aside from addressing the previously mentioned shortcomings, it would be interesting to analyze cases of data augmentation involving other forms of data collection, such as those used in VAAs. Another interesting avenue of research is that of aligning LLMs toward different political-viewpoints and then summarizing these views using other LLMs, as shown in the work of [77]. Furthermore, it would be interesting to explore if augmentation is improved by RAFT ([78]), a recent proposal based on combining additional external knowledge (RAG) and fine-tuning. Ultimately, these advancements will require further interdisciplinary research as the potential impact of AI as a tool for augmenting democracy is a rather uncharted territory. We hope these findings contribute to stimulate and organize that exploration.

## Acknowledgments

We acknowledge the support of the Agence Nationale de la Recherche grant number ANR-19-P3IA-0004, the European Union *LearnData, GA no. 101086712* a.k.a. 101086712-LearnData-HORIZON-WIDERA-2022-TALENTS-01 (https://cordis.europa.eu/project/id/101086712), IAST funding from the French National Research Agency (ANR) under grant ANR-17-EURE-0010 (Investissements d'Avenir program), the European Lighthouse of AI for Sustainability [grant number 101120237-HOR-IZON-CL4-2022-HUMAN-02], the Obs4SeaClim 101136548-HORIZON-CL6-2023-CLIMATE-01 and ANITI ANR-19-P3IA-0004.

# Supplementary material

Supplementary material was submitted together with the manuscript.

### Ethics
This study did not involve original human subject's research.

### Data Accessibility
No primary data was collected during this study.

### Authors' Contributions
JFG, UG and CAH designed the study. JFG performed all calculations. JFG and CAH created the figures. CAH wrote the paper.

### Competing Interests
We declare no competing interests.

# Large Language Models (LLMs) as Agents for Augmented Democracy: Supplementary Material

## Jairo F. Gudiño[1], Umberto Grandi[2], and César Hidalgo[1,3]

[1] *Center for Collective Learning, University of Toulouse & Corvinus University of Budapest*
[2] *IRIT, Université Toulouse Capitole, Toulouse, France*
[3] *IAST, Toulouse School of Economics and University of Toulouse Capitole*

**APPENDIX A: Dataset and Proposals Description**

A total number of 67 policy proposals presented in Brazucracia.org were collected and curated from three presidential candidates in 2022 Brazil Presidential Elections (Luiz Inácio Lula da Silva, Ciro Gomes and Jair Bolsonaro) [1]. This curation work was led by a lawyer. The list of policy proposals is presented in Table 1. As asking to order preferences between 67 proposals is challenging, this experiment was divided into universes with 2 or 5 proposals. We focus on the case with 2 proposals as recent literature has found that LLMs struggle in predicting selections when 3 or more proposals are presented [2].

| Number | ID | Policy Proposal | Presidential Candidate |
|--------|----|-----------------|------------------------|
| 1 | 1 | Maintain current labor legislation | Bolsonaro |
| 2 | 2 | Write a new labor legislation to include modern labor regulations and social protection | Lula |
| 3 | 3 | Regulation and protection of workers' labor rights by application | Bolsonaro, Lula |
| 4 | 4 | Valorize the minimum salary to recuperate the purchasing power | Lula |
| 5 | 5 | Equal pay policy between men and women performing the same function | Bolsonaro, Lula |
| 6 | 6 | Creation of policies to provide hybrid work and home office for women with children | Bolsonaro |
| 7 | 7 | Creation of policies that guarantee the inclusion and permanence of the LGBTQIA+ population in the labor market | Lula |
| 8 | 8 | Expand, redesign, and improve the qualification programs of the police | Lula |
| 9 | 9 | Implement national guidelines for the promotion and defense of police human rights | Lula |
| 10 | 10 | Improve public job positions and salary plans with incentives related to goals | Bolsonaro |
| 11 | 11 | Commitment to the goals stipulated by the National Education Plan | Lula |
| 12 | 12 | Invest in specific programs and actions aimed at the educational recovery of those affected by the pandemic | Bolsonaro, Lula |
| 13 | 13 | Continue the policy of social and racial quotas for admission to higher education | Lula |
| 14 | 14 | Actions that guarantee internet access in public schools | Bolsonaro, Lula |
| 15 | 15 | Strengthening career plans and valuing teachers | Bolsonaro, Lula |
| 16 | 16 | Actions aimed at training and qualification of teachers | Bolsonaro |
| 17 | 17 | Establish the basic foundations of the subjects, removing ideological connotations and with a view to parents as the main actors in children's education | Bolsonaro |

| 18 | 18 | Strengthen democratic, secular and inclusive education with specific policies for people with disabilities, the LGBTQIA+ population and among other vulnerable groups | Lula |
|---|---|---|---|
| 19 | 19 | Investing in vocational education in line with labor market expectations | Ciro |
| 20 | 20 | Invest in the national system to promote technological development through funds and public agencies such as CNPq and CAPES | Lula |
| 21 | 21 | Combining face-to-face teaching with distance learning in basic education, analyzing regional peculiarities | Bolsonaro |
| 22 | 22 | Strengthening the national vaccination program | Lula |
| 23 | 23 | Strengthen the popular pharmacy program | Lula |
| 24 | 24 | Invest on the management of the SUS | Bolsonaro, Lula |
| 25 | 25 | Expand the articulation between the public and private health sectors | Ciro |
| 26 | 26 | Encourage research related to medicines | Ciro |
| 27 | 27 | Specific health policies aimed at women, LGBTQIA+ population, disabled people and among other vulnerable groups | Lula |
| 28 | 28 | Continue programs related to encouraging physical activity for primary care | Bolsonaro |
| 29 | 29 | Reinforce the consolidation of the national cancer care support program | Bolsonaro |
| 30 | 30 | Structuring the medical career in the SUS with mechanisms of attraction and recognition | Ciro |
| 31 | 31 | Health facilities with good performance should monitor and assist those with lower performance | Ciro |
| 32 | 32 | Expand the privatization of state-owned companies and national infrastructure concessions | Bolsonaro |
| 33 | 33 | Policy for valuing state-owned companies and those against privatization | Lula |
| 34 | 34 | Actions to encourage the creative economy | Lula |
| 35 | 35 | Policies and actions for debt renegotiation of households and companies | Lula |
| 36 | 36 | Revocation of the spending ceiling | Lula |
| 37 | 37 | Tax reform with change in burden reducing taxation on consumption and increasing income progressively so that the richest pay more | Lula |
| 38 | 38 | No income tax for workers making up to 5 minimum wages. | Bolsonaro |
| 39 | 39 | Actions to curb tax evasion | Lula |
| 40 | 40 | New fuel pricing policy | Lula |
| 41 | 41 | Consolidate and expand land regularization actions, allied to the strengthening of legal institutions that ensure access to firearms | Bolsonaro |
| 42 | 42 | Encouraging female entrepreneurship through the facilitation of credit and microcredit | Bolsonaro |
| 43 | 43 | Encouraging entrepreneurship through credit facilitation and debureaucratization | Bolsonaro, Lula |
| 44 | 44 | Reduce agricultural production costs and marketing price | Lula |
| 45 | 45 | Create a new National LGBTI+ Public Policy Committee | Ciro |
| 46 | 46 | Increase the participation of women in politics and public management | Bolsonaro, Lula |
| 47 | 47 | The complete opening of banking and fiscal secrecy of first and second level positions in the Executive Power. | Ciro |
| 48 | 48 | Country's formal adherence to the OECD Council's Public Integrity Recommendation | Bolsonaro |
| 49 | 49 | Implement a federal Public Integrity strategy | Bolsonaro |
| 50 | 50 | Propose rules for the transparency of final beneficiaries of public resources | Bolsonaro |
| 51 | 51 | Investment in the Armed Forces and promotion of their international participation as in UN-sponsored missions | Bolsonaro |
| 52 | 52 | Maintain the value of 600 reais for Auxílio Brasil | Bolsonaro |
| 53 | 53 | Create a program that expands the guarantee of citizenship for the most vulnerable and brings a universal minimum income | Lula |

| 54 | 66 | Culture-focused policies through articulation with private sector institutions and companies and civil society organizations | Bolsonaro |
| 55 | 67 | Increase transparency through compliance with the Access to Information Law | Lula |
| 56 | 68 | Preservation of culture and demarcation of indigenous and quilombolas lands | Lula |
| 57 | 54 | Encouraging mining activity within a logic of environmental protection | Bolsonaro, Lula |
| 58 | 55 | Actions to combat illegal mining | Bolsonaro, Lula |
| 59 | 56 | Increase national production of fertilizers | Bolsonaro, Lula |
| 60 | 57 | Improve and reduce the prices of transport services through the structuring of concessions and public-private partnerships | Bolsonaro |
| 61 | 58 | Encouraging sustainable agricultural practices | Bolsonaro, Lula |
| 62 | 59 | Fostering agro-industry and national production of inputs | Bolsonaro, Lula |
| 63 | 60 | Strengthen the energy supply with the expansion of clean and renewable sources | Bolsonaro, Lula |
| 64 | 61 | Offer Green Bonds to finance investments considered sustainable in the areas of transport, energy and between others | Bolsonaro |
| 65 | 62 | Meet the carbon gas reduction targets assumed by the country at the 2015 Paris Conference | Lula |
| 66 | 63 | Recover lands deteriorated by predatory activities and reforestation of devastated areas | Bolsonaro, Lula |
| 67 | 65 | Curb drug mining and money laundering in the Amazon by increasing the number of ecological bases | Bolsonaro |

Table 1. Policy proposals presented in Brazucracia. Source: [1]

The two-proposals universe was displayed as a classical pairwise comparison screen. Participation in the platform was anonymous.

In Table 2 we illustrate all possible values for each feature that is added as context in the prompt.

| Feature | Description |
| --- | --- |
| Age | Young, Old (top and bottom quartiles) |
| Political Ideology | Conservative, Centrist, Liberal |
| Zone | Urban, Rural |
| Educational Attainment | College educated, non-college educated |
| Sex | Male, Female |
| Location | City and State |

Table 2. Self-reported demographic characteristics of participants in Brazucracia. Source: [1]

## APPENDIX B: Data adequacy checks

To check if we possess enough data to make a fair assessment of the predictive power of LLMs, we compute the smallest number of participants in a random sample whose aggregated preferences closely resemble those of another random sample of the same size. If both random samples are small, we have adequate data as aggregated preferences show small variability. Following this intuition, we take two non-overlapping random samples with equal number of participants and calculate the coefficient of determination ($R^2$) of the Pearson correlation of the win rates of each proposal between both (the definition of win-rates is presented in the main document). Plotting the $R^2$ value against the number of participants in each random sample according to Figure 1, we estimate that—for instance—a minimum of 53 participants (20% of the sample) is required to attain a $R^2$ equal to 0.72. We estimate the test-retest consistency of our sample (X=100%) to be $R^2$=95.38%.

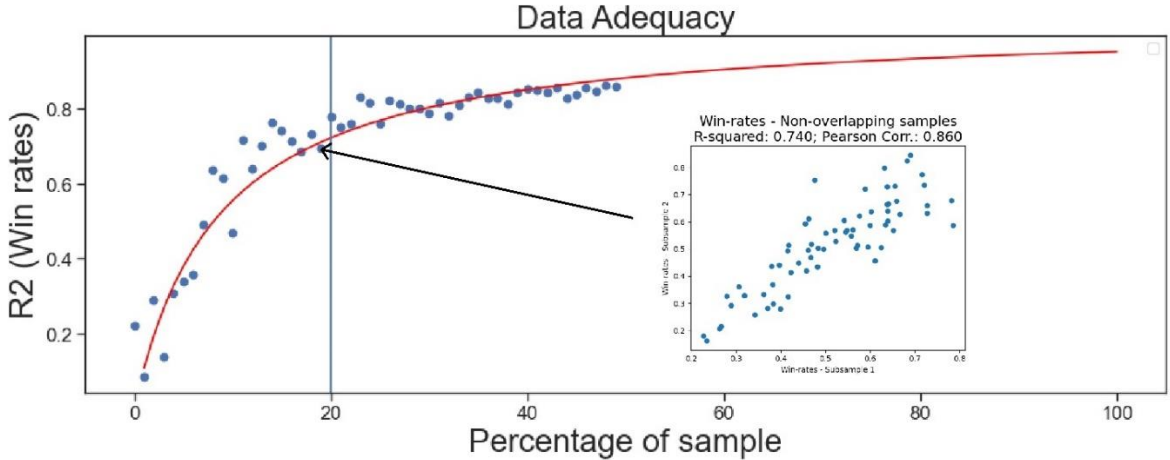

Figure 1. Data adequacy assessment based on comparing aggregations between two non-overlapping groups of participants as function of the size of each sample (as percentage of the population). When taking two non-overlapping samples of 53 participants (20% of the population) the R-squared of the Pearson correlation is 72.2%. A rational function is used to calculate the trend depicted in red color.

**APPENDIX C: Fine-tuning LLMs, Code and Repositories**

In this appendix, we present details about processes to fine-tune different LLMs and the repository of code and models used to produce results. We used the Python libraries *transformers*, *peft*, *trl*, *bitsandbytes* and *torch* for downloading, compressing, training, and saving models in HugginFace. We use the library vLLM library to improve the throughput of LLMs ([3]). To ensure comparability of results, the model architecture of each LLM is the same across different sizes of the training set.

**LLMs' hyperparameters**

*BERT SQuAD:*

As a foundational LLM, its predictive accuracy relies heavily on the quality of embeddings we provide for contextual understanding. To ensure this, all prompts were initially translated into Brazilian Portuguese using the prompt presented in Appendix C, enabling us to leverage BERTimbau embeddings (publicly available in HuggingFace in "neuralmind/bert-base-portuguese-cased") as our starting point for the training [4]. These embeddings were derived by using brWaC, the largest corpus to date in this language comprising 3.53 million documents (2.68 billion tokens). To fine-tune a BERT model in a Question Answering problem (multiple choice), we select as optimal hyperparameters 2 epochs, a learning rate of 0.00005 and a weight decay rate equal to 0.01. Training was conducted on a single CPU with 50 GB of memory.

*GPT 3.5 Turbo:*

As GPT 3. 5 Turbo is based on proprietary neural architecture, we use the OpenAI Python API to fine-tune GPT 3.5 models ("gpt-3.5-turbo"), fixing the number of epochs to 3 as the only relevant hyperparameter and letting GPT automatically decides the batch size and learning rate.

*Mistral 7B:*

We fine-tuned the second version of Mistral Instruct with 7 billion of parameters ("mistralai/Mistral-7B-Instruct-v0.2"). To adapt this pre-trained LLM to our downstream application, we use Low-Ranking Adaptation of LLMs (LoRA) by setting the rank of decomposition (r) to 64, the scaling parameter α to

16 and the dropout probability to 0.1. For fine-tuning we set the optimal hyperparameters to 2 epochs, a learning rate of 0.0002, a weight decay rate equal to 0.001, a maximum gradient norm equal 0.3, the proportion of training steps to use for warming up the learning rate as 0.03 and a constant learning rate scheduler. The maximum number of steps is set to 8,250. We train this LLM in one T4 GPU with 16-bit precision.

### LLaMA-2 7B & Falcon 7B:

We fine-tuned a second version of LLaMA with 7 billion of parameters for chat ("NousResearch/Llama-2-7b-chat-hf") and an instruct version of Falcon with 7 billion of parameters ("vilsonrodrigues/falcon-7b-instruct-sharded") with the same hyperparameters. To adapt these pre-trained LLMs to our downstream application, we use LoRA by setting the rank of decomposition (r) to 64, the scaling parameter $\alpha$ to 16 and the dropout probability to 0.1. For fine-tuning we set the optimal hyperparameters to 3 epochs, a learning rate of 0.0002, a weight decay rate equal to 0.001, a maximum gradient norm equal 0.3, the proportion of training steps to use for warming up the learning rate as 0.03 and a cosine learning rate scheduler. The maximum number of steps is set to 3,000. We train these LLMs in one T4 GPU with 16-bit precision.

For predicting policy preferences of the population, we set the temperature to 0 as inference parameter for all LLMs to ensure reproducibility of results.

## Code Repository

The code used to reproduce results will be posted in GitHub. The link will be updated in this section.

## Models Repositories

Fine-tuned LLMs used in this are publicly available in the GitHub repo.

**APPENDIX D: Prompts in Brazilian Portuguese**

For robustness, we also created prompts in Brazilian Portuguese that encapsulate the demographic characteristics of each participant. In Figure 2 we present a table of the possible values of demographic characteristics in this language and reproduce the Figure 1 b of the main document as an example for a 26 years-old liberal male with a master's degree living in Rio de Janeiro, who in *Brazucracia* selected the proposal: "actions to curb tax evasion" over the proposal "expand the privatization of state-owned companies and national infrastructure concessions".

| Feature | Description (in Brazilian Portuguese) |
|---|---|
| Age | Jovem, De terceira idade |
| Political Ideology | Conservador, Centrista, Esquerdista |
| Zone | Urbana, Rural |
| Educational Attaintment | Sem curso superior, Com curso superior |
| Sex | Masculino, Feminino |
| Location | [City and State] |

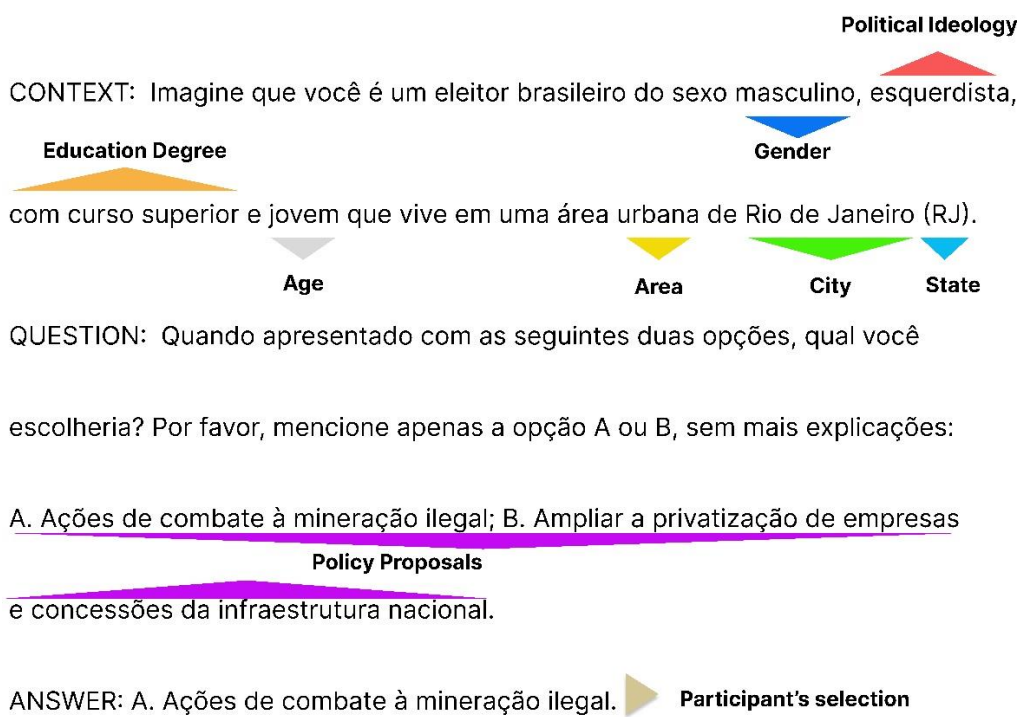

**Figure 2. a** Table of features and values in Brazilian Portuguese. **b** Example of the "mad-libs" style prompt in Brazilian Portuguese used to fine-tune the LLMs.

## APPENDIX E: Accuracy of LLMs

For robustness, we reproduce Figure 1 d (in the main document) for different sizes of random samples from the population (5%, 25% and 50% of participants). Results are presented in Figure 3.

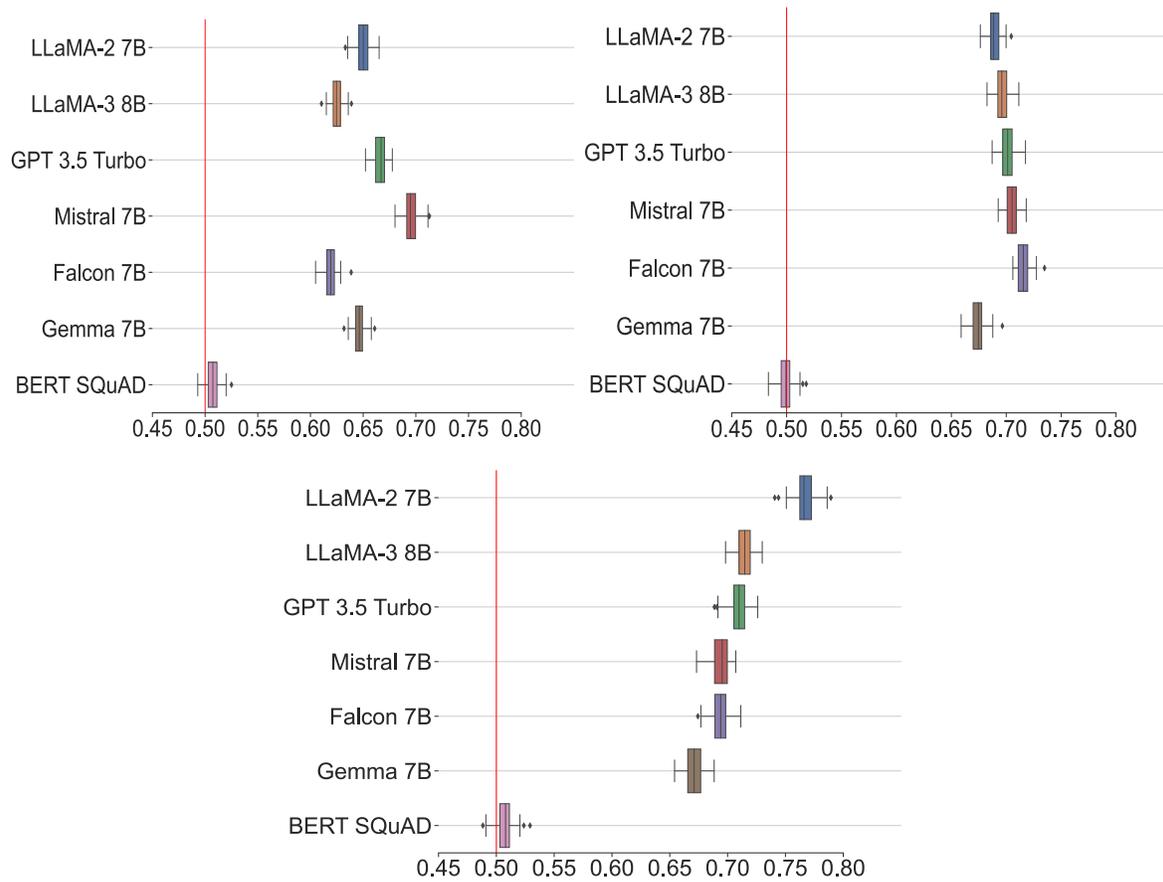

Figure 3. Accuracies obtained on 5%, 25% and 50% test set of five LLMs.

# APPENDIX F: Kendall Rank Correlation Coefficient of LLMs

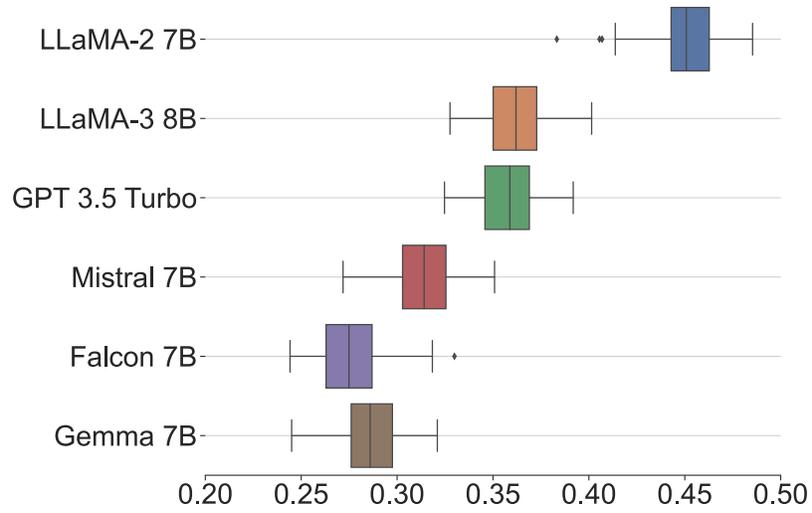

Figure 4. Kendall Rank Correlation obtained on the 50% test set of six LLMs.

**APPENDIX G: Consistency rate of LLMs**

For each prompt (containing demographic characteristics and the pair of proposals) the answer of an LLM can be "A", "B" or an inconsistent reply as the answer depends on the order that options are written (A and then B, or B and then A). To analyze how frequent this last case is, we compute the "*consistency rate*", which is the number of times that a certain preference (A or B) is chosen by an LLM regardless of the order of the options divided by the total number of pairwise preferences. We present in Figure 5 the values of this metric across different LLMs and sizes of probabilistic samples. While GPT-3.5 Turbo and Falcon 7B deliver consistent replies in general (>70%), this is not the case for the rest despite similar values of accuracy.

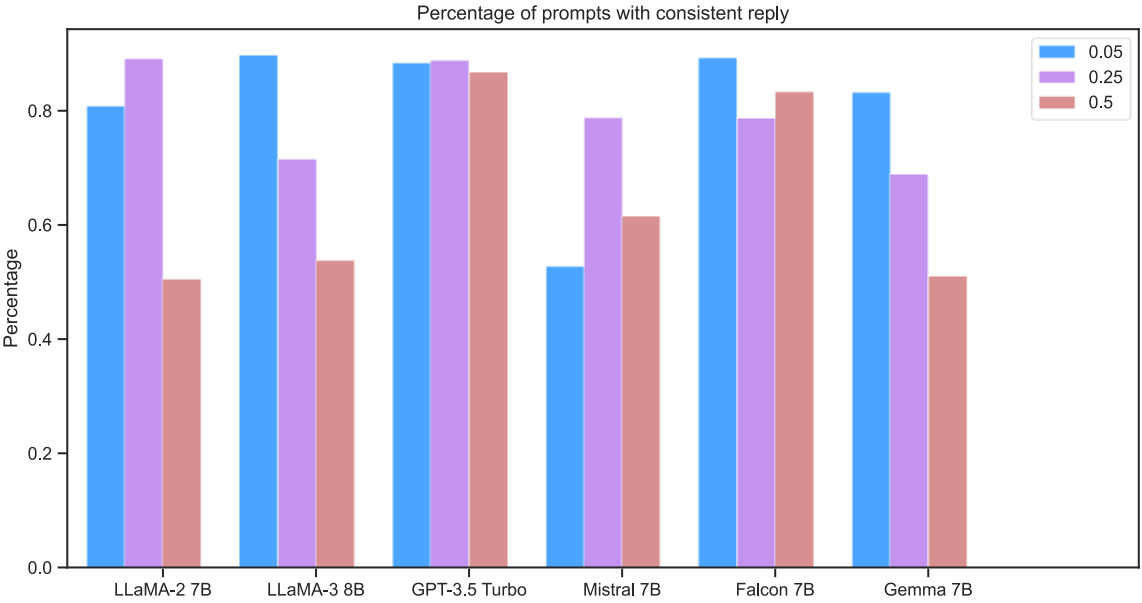

Figure 5. Percentage of LLMs predictions with consistent replies. Each color represents a probabilistic sample.

**APPENDIX H: Sensitivity tests for temperature**

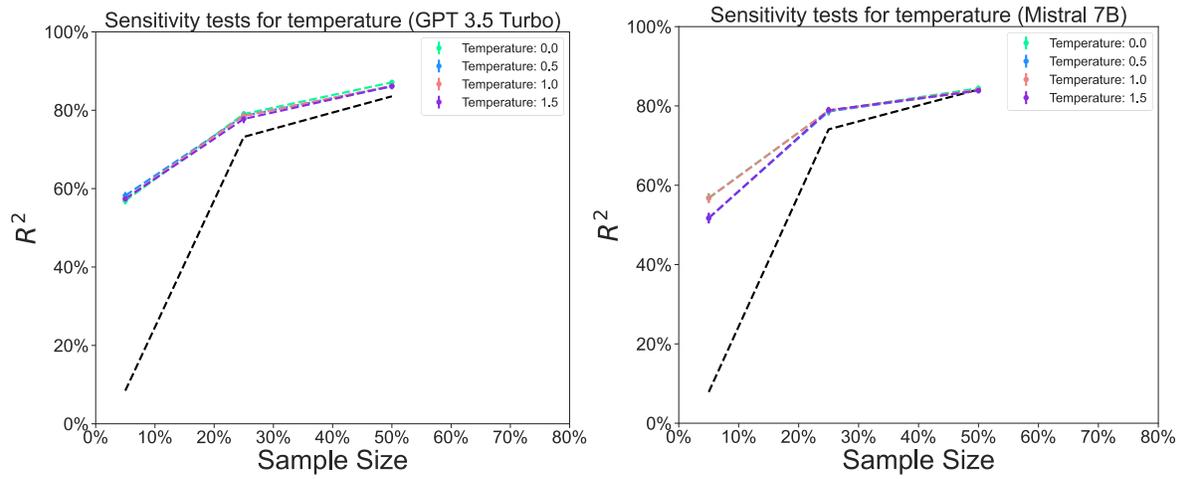

Figure 6. Sensitivity tests for GPT 3.5 Turbo and Mistral 7B.

## APPENDIX I: t-tests of equality of means for two samples

To establish if accuracy values are different across different pairs of subgroups by using the same LLM, we calculate p-values associated to t-tests of equality of means for two-samples with unequal variances [5]. The null hypothesis posits that in each LLM the mean accuracy in the first subgroup is equal to the accuracy in the second group. For instance, in the comparison between college-educated vs. non-college educated individuals, we assess whether college-educated participants exhibit similar accuracy in average compared to their non-college educated counterparts. The results of these pairwise comparisons are presented in Table 10.

| LLM | Comparison | t-Statistic | p-value |
|---|---|---|---|
| GPT 3.5 Turbo | College-educated vs. non-Collage ed. | -6.9619377 | 0*** |
| LLaMA-2 7B | College-educated vs. non-Collage ed. | -11.762422 | 0*** |
| Mistral 7B | College-educated vs. non-Collage ed. | -14.852711 | 0*** |
| Falcon 7B | College-educated vs. non-Collage ed. | -4.122438 | 0.0001*** |
| LLaMA-3 8B | College-educated vs. non-Collage ed. | -13.954066 | 0*** |
| Gemma 7B | College-educated vs. non-Collage ed. | -7.5638435 | 0*** |
| GPT 3.5 Turbo | Liberal vs. Conservative | -20.601809 | 0*** |
| GPT 3.5 Turbo | Liberal vs. Centrist | -3.8584591 | 0.0002*** |
| LLaMA-2 7B | Liberal vs. Conservative | -26.47836 | 0*** |
| LLaMA-2 7B | Liberal vs. Centrist | -10.614338 | 0*** |
| Mistral 7B | Liberal vs. Conservative | -30.800306 | 0*** |
| Mistral 7B | Liberal vs. Centrist | -8.6633955 | 0*** |
| Falcon 7B | Liberal vs. Conservative | -25.976078 | 0*** |
| Falcon 7B | Liberal vs. Centrist | -9.5934272 | 0*** |
| LLaMA-3 8B | Liberal vs. Conservative | -12.609069 | 0*** |
| LLaMA-3 8B | Liberal vs. Centrist | -4.8693856 | 0*** |
| Gemma 7B | Liberal vs. Conservative | -19.443508 | 0*** |
| Gemma 7B | Liberal vs. Centrist | -9.3729255 | 0*** |
| GPT 3.5 Turbo | Younger quartile vs. Older quartile | 0.5821314 | 0.5616 |
| LLaMA-2 7B | Younger quartile vs. Older quartile | 1.3343591 | 0.1842 |
| Mistral 7B | Younger quartile vs. Older quartile | -3.5098211 | 0.0006*** |
| Falcon 7B | Younger quartile vs. Older quartile | -3.7848422 | 0.0002*** |
| LLaMA-3 8B | Younger quartile vs. Older quartile | -1.2293032 | 0.221 |
| Gemma 7B | Younger quartile vs. Older quartile | -4.5272051 | 0*** |
| GPT 3.5 Turbo | Female vs. Male | 1.6997025 | 0.0908 |
| LLaMA-2 7B | Female vs. Male | -11.793089 | 0*** |
| Mistral 7B | Female vs. Male | -8.2075363 | 0*** |
| Falcon 7B | Female vs. Male | -7.2997403 | 0*** |

| LLaMA-3 8B | Female vs. Male | 3.9603965 | 0.0001*** |
| Gemma 7B | Female vs. Male | 5.5989315 | 0*** |

Table 10. T-statistics and p-values associated to t-tests for equality of means between two samples with unequal variances for different LLMs trained on half of the population (GPT 3-5 Turbo, Mistral 7B, LLaMA-2 7B, Falcon 7B, , LLaMA-3 8B and Gemma 7B). The null hypothesis posits that means are equal. *** denotes 1% statistically significant level, ** denotes 2% significance level, and * denotes 5% level.

# Bibliographical References